\documentclass[12pt]{article}
\newcommand{\mc}{\mathcal}
\newcommand{\dis}{\displaystyle}

\usepackage [dvips]{graphicx}
\usepackage[english]{babel}
\usepackage{amsfonts}
\usepackage{amsmath}

\title{\bf Option pricing and hedging with minimum local expected shortfall}
\author{Benoit Pochart \thanks{Centre de Math\'ematiques Appliqu\'ees,
Ecole Polytechnique, 91128 Palaiseau {\sc Cedex}, FRANCE (Email:
pochart@cmapx.polytechnique.fr).} \, and Jean-Philippe
Bouchaud\thanks{Service de Physique de l'Etat Condens\'e, Centre
d'\'etudes de Saclay, Orme des Merisiers, 91191 Gif-sur-Yvette
{\sc Cedex}, FRANCE (Email: bouchau@drecam.saclay.cea.fr).}
\thanks{Science \& Finance, Capital Fund Management,
109-111 rue Victor Hugo, 92532 Levallois {\sc Cedex}, FRANCE,
http://www.science-finance.fr. } }

\date{}
\begin{document}
\maketitle \abstract{We propose a versatile Monte-Carlo method for pricing and hedging
options when the market is incomplete, for an arbitrary risk criterion (chosen here to
be the expected shortfall), for a large class of stochastic processes, and in the presence of transaction costs.
We illustrate the method on plain vanilla options when the price returns follow
a Student-t distribution. We show that in the presence of fat-tails, our strategy allows to
significantly reduce extreme risks, and generically leads to low Gamma hedging.
Similarly, the inclusion of transaction costs reduces the Gamma of the
optimal strategy.}
\section{Introduction}

In their seminal 1973 article \cite{BS}, Black and Scholes (BS) have
founded the very basis of modern financial mathematics. Their work
has since been much studied and refined, and has become a rather abstract
conceptual framework, deeply related with modern Probability Theory
\cite{HP,Musiela}. The BS model is the paradigm of \emph{complete
markets}, where every contingent claims can be replicated by a
portfolio of underlying assets. Using a no-arbitrage principle one can deduce
that any option has a unique price, independent of the agent's risk
preferences, which is given by the price of the replicating (or hedging) strategy.
Mathematically speaking, these properties are equivalent to the
existence of a unique equivalent martingale measure (also called the risk-neutral measure)
under which one should average the final pay-off of an option to obtain its price.
This measure is in general different from the `true' (or objective) real world probability measure
(under which we observe the evolution of financial assets) \cite{HP,Musiela}.
The knowledge of this true probability distribution is thus in principle
of no use for the pricing of options, although, as discussed in detail in \cite{PBS,Potters},
this message is in fact rather misleading. Within the BS framework, analytical formulae
for the price and hedge exist for several cases, such as the simple European
options. On the other hand, for more complex products like American or
path dependent options, numerical procedures often have to be used \cite{Hull,Wilmott}.

However, the hypotheses of the BS model (Gaussian distribution and
independence of log-returns, continuous time, absence of friction,
etc.) have been widely questioned by practitioners and one can now
observe a growing interest in the academic community for more
general stochastic processes (L\'evy processes
\cite{Potters,Levendorskii,GMY,Eberlein}, stochastic volatility
\cite{Stein,Heston,Yakov,Masoliver,Masoliver2}, multifractal
processes \cite{Muzy,Calvet,Pochart}) which usually result in much
more complex \emph{incomplete markets}. In this case contingent
claims cannot in general be replicated using underlying assets and
the no-arbitrage principle is no longer enough to yield a unique
price. One has to introduce additional criteria related to the
risk preferences of the agents: the absence of perfect replication
implies the existence of an intrinsic residual risk
\cite{Potters}.

In this framework one simple and popular methodology is the
variance-minimizing hedging strategy \cite{Schweizer,Potters}. It
consists in finding the (self-financing) portfolio whose
difference with the pay-off of the option at maturity has the
smallest variance. One drawback of this approach is that the
risk-function is quadratic and therefore penalizes both profits
and losses. Also, a quadratic measure of risk does not strongly
penalize extreme risks. Alternative criteria based on higher
moments of the distribution or on the notion of Value at Risk
(VaR) and its extensions \cite{Delbaen} have been recently
considered in the literature \cite{Follmer,Uryasev,Selmi,Potters}.
In \cite{Follmer} the authors are interested in finding, given an
initial investment, the portfolio strategy which maximizes the
probability of a successful hedge. Treated in a general setting
and in a rather abstract way, this problem is shown to reduce to
the replication of a particular knockout option. Although very
appealing this solution can be hard to implement from a practical
point of view. Indeed an explicit form of the option is only
available in specific models and even in that case, as highlighted
by the authors, the practical replication of such an option is not
an easy task. The approach of \cite{Uryasev} is much more concrete
and numerically oriented. Nevertheless it is unsatisfactory from
an optimization point of view: it is a static approach (the
weights are determined at the initial time and remain constant)
not well adapted to determine the full dynamical replication
strategy. A similar observation can be made concerning the work
presented in \cite{Selmi}, where the optimal static strategy that
minimizes the fourth moment of the profit and loss distribution is
determined. The approximate dynamical strategy is then constructed
by `translating' (in time) the optimal static strategy. An
interesting observation made in that work is that the hedging
strategy varies less rapidly with the underlying than the
quadratic hedge, implying lower transaction costs (see also
\cite{Potters}).

The aim of this paper is to propose a general Monte-Carlo algorithm
that allows to price and hedge options using an arbitrary (but sufficiently smooth)
risk criterion, and for a large class of stochastic processes describing the
underlying. The
algorithm, that generalizes the work of \cite{PBS}, is both easy to implement and versatile,
and can be used for pricing different types of options. The paper is organized as
follows: we first explain, following \cite{PBS}, the main ideas of our
methodology. We then present some numerical results for different
underlying processes, which show that extreme risks can be efficiently
reduced compared to the standard BS hedge. As in \cite{Selmi}, we find that these
extreme hedges have a smaller `Gamma' at and around the money. Finally, we show how
transaction costs can be treated within our method, and discuss several other possible
extensions of our scheme.

\section{Description of the method}

\subsection{Notations}

For simplicity we consider the case of a European option with one
underlying asset of maturity $\dis{T=N \tau}$, where $N$ is the
number of re-balancing dates and $\tau$ the time interval two
dates. We denote the price of the underlying asset at time
$\dis{t_k=k \tau}$ by $x_k$, the strike by $K$ and the final
pay-off is $(x_N-K)_+ \equiv \max(x_N-K,0)$. We suppose that the
price of the option only depends on the current price $x_k$ of the
asset and call it by $\mc{C}_k(x_k)$ at time $t_k$. [If the
volatility was stochastic, we should assume that the option price
also depends on the current level of volatility $\sigma_k$ and
rather write $\mc{C}_k(x_k,\sigma_k)$]. The interest rate is
assumed to be constant and equal to $r$ and we define $\rho=r
\tau$. Averaging (denoted by angled brackets $\langle ...
\rangle$) will in the following always refer to the {\it
objective} (real world) probability measure under which we observe
the distribution of the asset returns, and {\it not} any abstract
risk neutral measure.

\subsection{Principles}

The method we investigate here is an extension of the hedged
Monte-Carlo strategy presented in \cite{PBS}. Our aim is to
construct a self-financing portfolio, whose wealth variation only
depends on the variation of the asset price \cite{Musiela}, that
best minimizes the chosen (instantaneous) risk measure. We denote
by $\phi_k(x_k)$ the fraction of the underlying asset in the
portfolio at time $k$, when the asset price is $x_k$. Between time
$t_k$ and $t_{k+1}$ the self-financing condition leads to a local
wealth balance given by: \cite{PBS}
\begin{equation}\label{balance}
   \Delta W_k=e^{\rho}\mc{C}_k(x_k)-\mc{C}_{k+1}(x_{k+1})+\phi_k(x_k)
   (x_{k+1}-e^{\rho} x_k)
\end{equation}
The measure of the quality of the replication is given by a local risk function
${\cal U}(\Delta W_k)$. The average risk, over all paths of the real process, is thus
given by:
\begin{equation}\label{risk}
\mc{R}_k= \big \langle  {\cal U}(\Delta W_k) \big \rangle.
\end{equation}
For purposes of illustration, we have chosen in the following a
function ${\cal U}(\Delta W_k)$ which penalizes losses that exceed
a certain threshold $-\Delta_0$:
\begin{equation}\label{risk2}
 {\cal U}(\Delta W_k) = (\Delta_0-\Delta W_k)^q_+ = |\Delta_0-\Delta W_k|^q
 \mathbf{1}_{\Delta W_k<\Delta_0},
\end{equation}
where the exponent $q$ was chosen to $q=1$, corresponding to what
is called an unconditional expected shortfall. The
generalization to arbitrary $q$, or in fact to other functional
forms for ${\cal U}(\Delta W_k)$, does not lead to any numerical
difficulty ($q > 1$ penalizes more strongly extreme losses). The
minimization of $\mc{R}_k$ is quite sensible from a financial
point of view: one tries to control in a marked to market way,
during the whole life of the option, the occurrence of downside
moves. Choosing a large negative $\Delta_0$ means that one aims at
controlling extreme losses.

From a practical point of view we solve the above optimization problem using
Monte Carlo simulations, that allows one to use a rather general stochastic
process for the price evolution. We generate $N_{MC}$ trajectories of the
asset price over which we will average. Following \cite{Longstaff,PBS},
we decompose the functions $\mc{C}_k(x)$ and $\phi_k (x)$ on a set of $p$
fixed basis functions:
\begin{eqnarray}
  \mc{C}_k(x) &=& \sum_{a=1}^{p} \gamma_a^k C_a^k(x) \\
  \phi_k (x) &=& \sum_{a=1}^{p} \varphi_a^k F_a^k(x)
\end{eqnarray}
Doing this we reduce the original functional optimization (find
the functions $\phi_k$ and $\mc{C}_k$) to a numerical optimization:
we now have a minimization problem in term of the parameters $\gamma_a^k$ and
$\varphi_a^k $. If $p$ is large enough we expect to
have a good approximation of the true functional solution (see the following
section for numerical implementation). We solve the problem by
working backward in time from maturity, where the option is
worth its known final pay-off. For each time $k$, we decompose the problem into the
following steps:
\begin{itemize}
    \item If the time discretisation mesh is sufficiently small, one can
    approximate $\mc{C}_{k}(x_{k})$ by
    $\mc{C}_{k+1}(x_{k})$ whose functional form is already known from the previous
    step.\footnote{A better approximation that takes into account the (already known) time
        derivative of $\mc{C}_{k}(x_{k})$ is to write $\mc{C}_{k}(x_{k}) \simeq 2\mc{C}_{k+1}(x_{k})
        -\mc{C}_{k+2}(x_{k})$.}
        We then find the coefficients $\varphi_a^k $ which minimize
    the average risk over the $N_{MC}$ paths:
    \begin{equation*}
    \mc{R}_k^* = \sum_{\ell=1}^{N_{MC}}
    \bigg (\Delta_0 -\left (e^{\rho}C_{k+1}(x_k^\ell)-C_{k+1}(x_{k+1}^\ell)\right)-
    (x_{k+1}^\ell-e^{\rho} x_k^\ell) \sum_{a=1}^{p}\varphi_a^k F_a^k(x_k^\ell)
    \bigg)_+.
    \end{equation*}
    This can be done using a steepest gradient method. Indeed we can easily compute
    the partial derivative of $\mc{R}_k^*$ with
    respect to the coefficients $\varphi_a^k $:
    \begin{equation*}
    \frac{\partial \mc{R}_k^*}{\partial \varphi_a^k}= -\sum_{\ell=1}^{N_{MC}}
    (x_{k+1}^\ell-e^{\rho} x_k^\ell) F_a^k(x_k^\ell)  \mathbf{1}_{\Delta W_k^\ell <\Delta_0}.
    \end{equation*}
    \item Using the fact that, on average, the local wealth balance must equal zero,
    we now compute the coefficients $\gamma_a^k$ by solving
    the least square problem:
    \begin{equation*}
    \min_{\gamma} \,\,\, \sum_{\ell=1}^{N_{MC}}\left [\sum_{a=1}^{p} \gamma_a^k C_a^k(x_k^\ell)
    -e^{-\rho}\bigg (\mc{C}_{k+1}(x_{k+1}^\ell)-\phi_k(x_k^\ell)(x_{k+1}^\ell-e^{\rho}
    x_k^\ell)\bigg) \right]^2,
    \end{equation*}
    which is easily done using standard procedures \cite{Numerical}.
\end{itemize}

\section{Numerical results}
\subsection{Implementation issues}
In this section we compare the results obtained following a
standard Black and Scholes strategy (delta hedging) with those
obtained following strategies with different values of the
threshold $\Delta_0$, as explained above. We price a European
option, with a maturity of $1$ year and an annualized volatility
$\sigma= 20\%$. We choose a rather small number of time intervals
when re-hedging is possible, $N=10$. The initial stock price is
$x_0=100$. We use $N_{MC}=20 000$ trajectories for averaging. We
first consider realizations of a standard geometric Brownian
motion with a constant drift $\mu=0.05$:
$$dx_t= x_t \left(\mu dt+ \sigma dW_t \right)$$
It is well known that financial time series are very poorly
represented by such a process and display much heavier tails. To
qualitatively account for this fact we also use realizations of a
fat tailed process where we replaced the previous Brownian motion
$W_t$ by a fat tailed process $L_t$ whose increments are
distributed according a Student$-t$ distribution, with $\nu=4$ or $\nu=6$
degrees of freedom:\footnote{See \cite{Lisa} for option pricing with
Student$-t$ distribution of returns on all time scales, to be contrasted
with the present model where only the shortest time scale returns are
distributed according to a Student$-t$ distribution.}
$$dx_t= x_t \left(\mu dt+ \sigma dL_t \right)$$
The value of $\nu$ characterizes the power-law decay of the
distribution for large arguments; $\nu=4$ is in the range of
reported value for this exponent for rather liquid markets. The
value $\nu=6$ corresponds to faster decaying tails; the limit $\nu \to\infty$
corresponds to the usual Black-Scholes model. Some
markets (like emerging country markets, or emerging country
currencies) would correspond to small values of $\nu$ (for example
$\nu \approx 1.5$ for the Mexican Peso). We use $p=20$ basis
functions, which we find to be accurate enough. Following
\cite{PBS} we choose for $F_a^k$ piecewise linear functions and
for $C_a^k$ piecewise quadratic functions with both the same
adaptive breakpoints. These breakpoints are chosen so that at each
stage the same number of simulated trajectories, $N_{MC}/(p+2)$,
fall between two successive breakpoints. $F_a^k$ is worth $0$
below the $a^{th}$ breakpoint, $1$ above the $a+1^{th}$ breakpoint
and is linear between these two values. $C_a^k$ is taken as the
integral of $F_a^k$ which is worth $0$ below the $a^{th}$
breakpoint. Our numerical simulations were systematically
conducted as follows: we first generate a set of trajectories and
apply our algorithm to find the coefficients $\gamma_a^k$ and
$\varphi_a^k$, i.e., the price of the option and the optimal
hedge. We then simulate a new set of paths to compute different
statistical indicators of the performance of the proposed
strategies. In other words, different paths are used for the
optimization, and for backtesting the optimization in an `out of
sample' fashion.

\subsection{Expected shortfall hedging in the Black-Scholes case}

\begin{figure}[!htbp]
  \begin{center}
    \includegraphics[angle=0,scale=0.55]{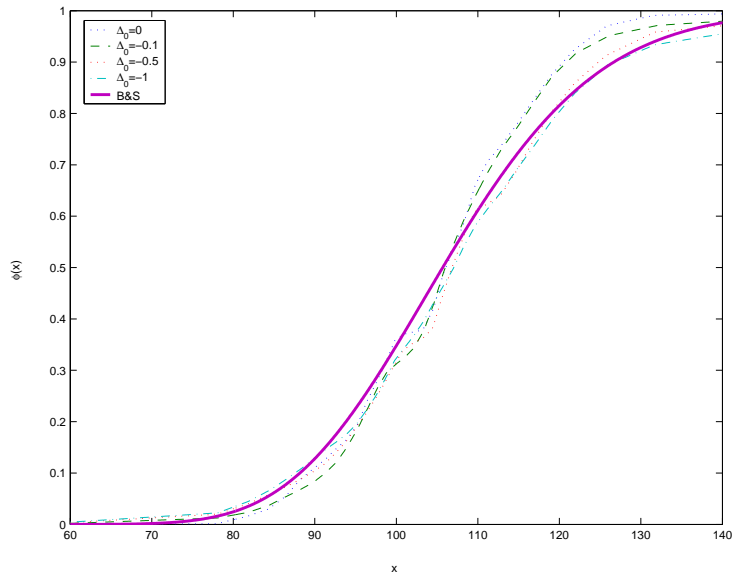}
    \caption{\label{fig} Optimal number of risky assets $\phi$ in the hedging portfolio,
    as a function of the level $x$ of the underlying asset for different strategies
    within the Black-Scholes. As expected the different curves are very similar.
    The value of the strike is $K=110$, and the time is half the maturity of the
        option: $k=N/2$.}
  \end{center}
\end{figure}

We first present the results obtained within the framework of the
Black and Scholes model. Since perfect replication is
theoretically possible in this case, we expect that the minimum
short-fall strategy we find should be close to the Black and
Scholes strategy. In the continuous time limit, the Black-Scholes
strategy actually leads to a zero expected shortfall, for any
value of $\Delta_0 < 0$. This is indeed what we observe in Fig.
\ref{fig} where we plot, for a strike price equal to $K=110$, the
optimal solutions found with different values of $\Delta_0$ and
the BS strategy: they all look very similar, in particular for
small values of $\Delta_0$ ($-0.5$ and $-1$).

\begin{figure}[!htbp]
  \begin{center}
    \includegraphics[angle=0,scale=0.4]{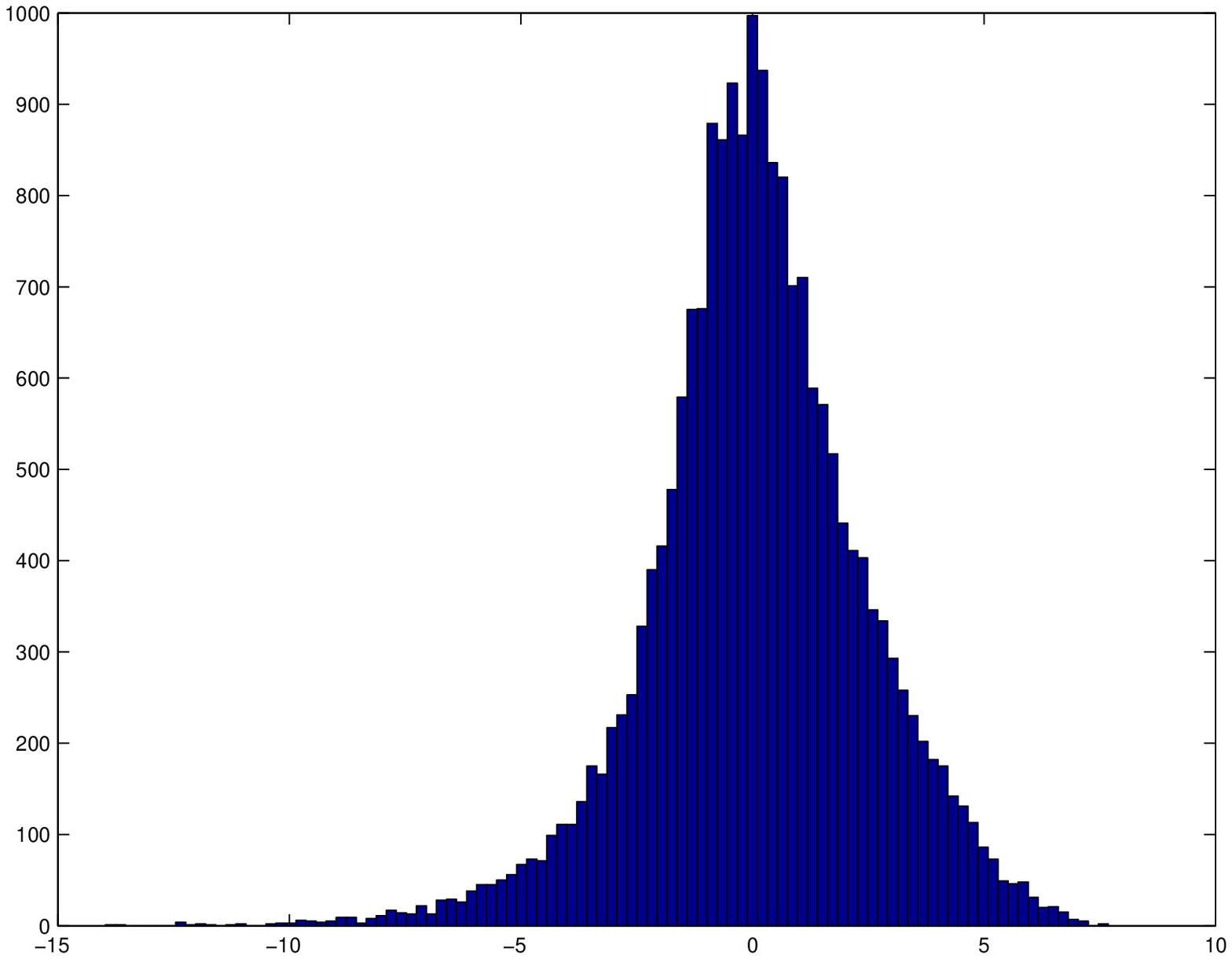}\includegraphics[angle=0,scale=0.4]{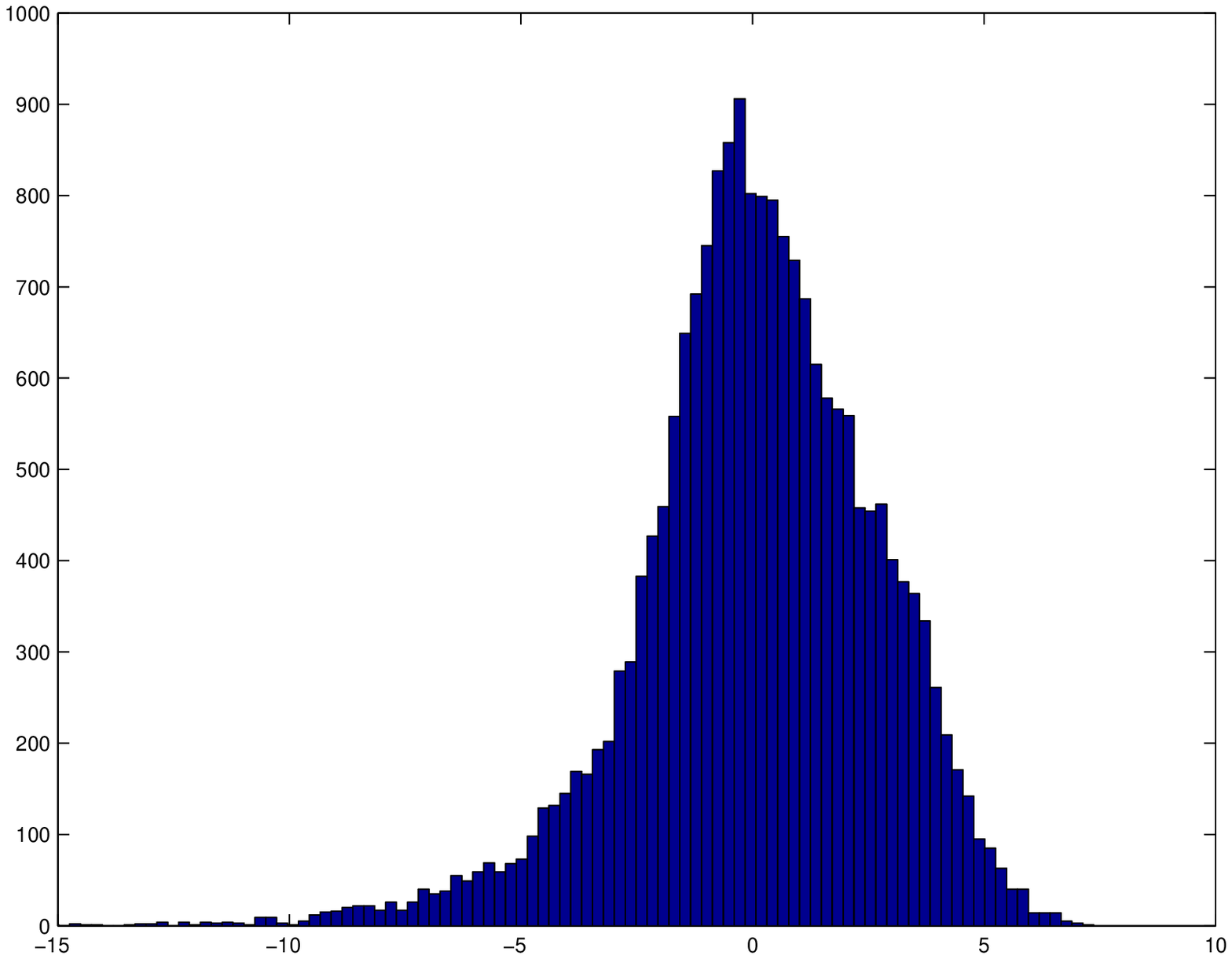}
    \includegraphics[angle=0,scale=0.4]{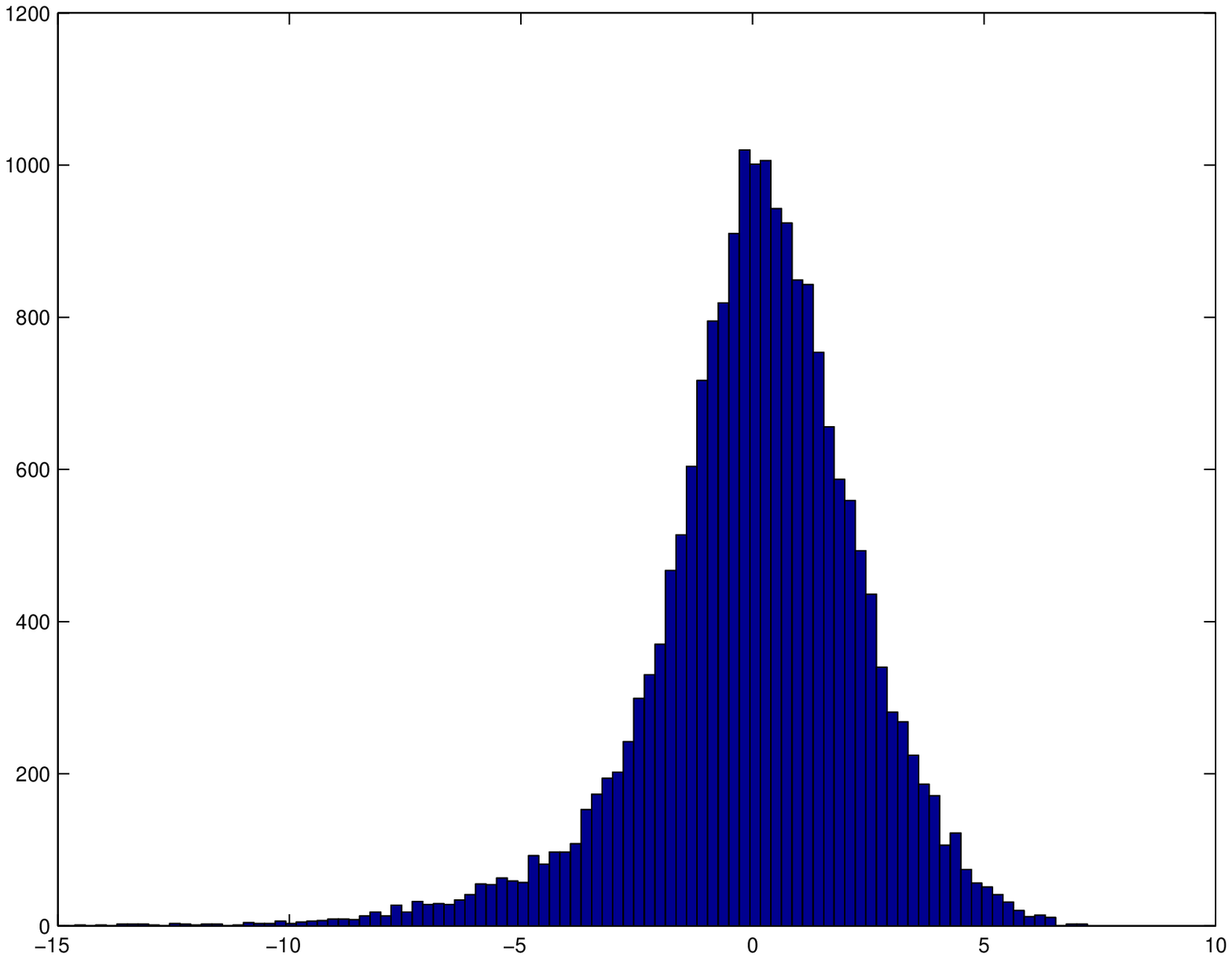}\includegraphics[angle=0,scale=0.4]{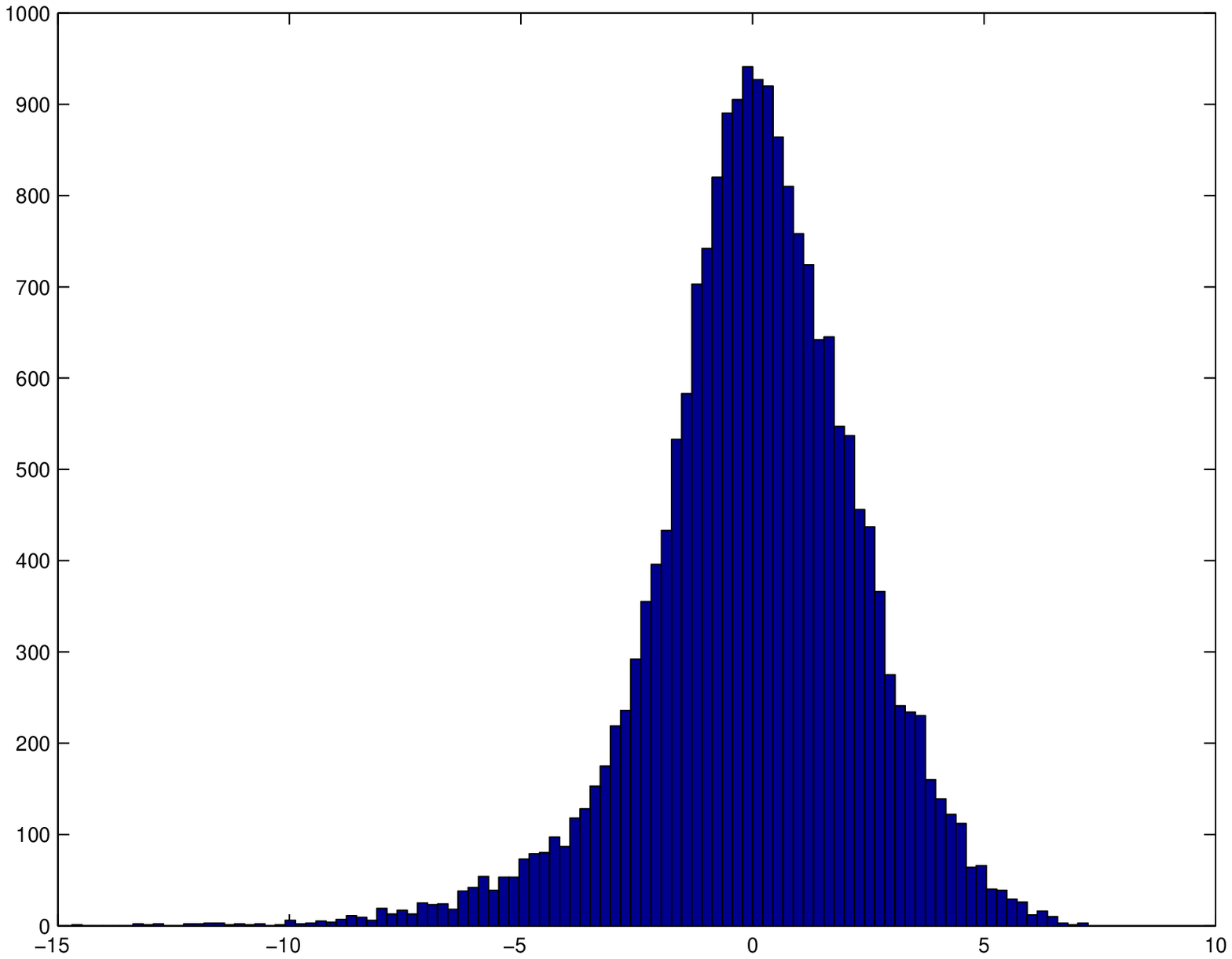}
    \caption{\label{dg} Distribution of the final wealth for different strategies for a
        BS market
    (upper left: BS, upper right: $\Delta_0=0$, lower left: $\Delta_0=-0.5$,
    lower right: $\Delta_0=-1$). The four curves are not very different from one another.
        The value of the strike is $K=110$.}
  \end{center}
\end{figure}

This observation is confirmed by the shape of the distribution of the
final wealth at the maturity date of the option, for the
different strategies (Fig. \ref{dg}).

\subsection{The case of a fat-tailed dynamics}

\begin{figure}[!htbp]
  \begin{center}
    \includegraphics[angle=270,scale=0.35]{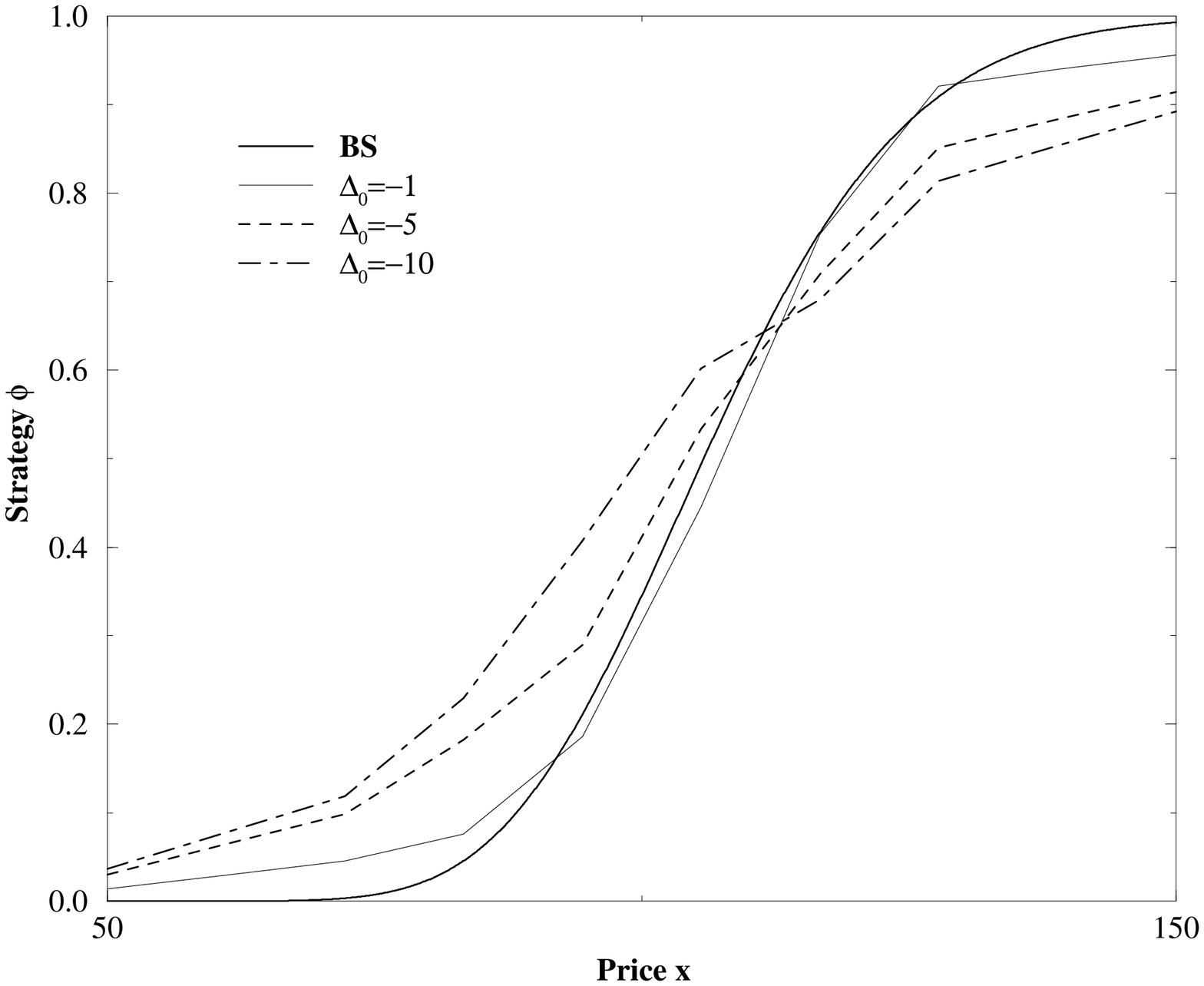} \includegraphics[angle=270,scale=0.35]{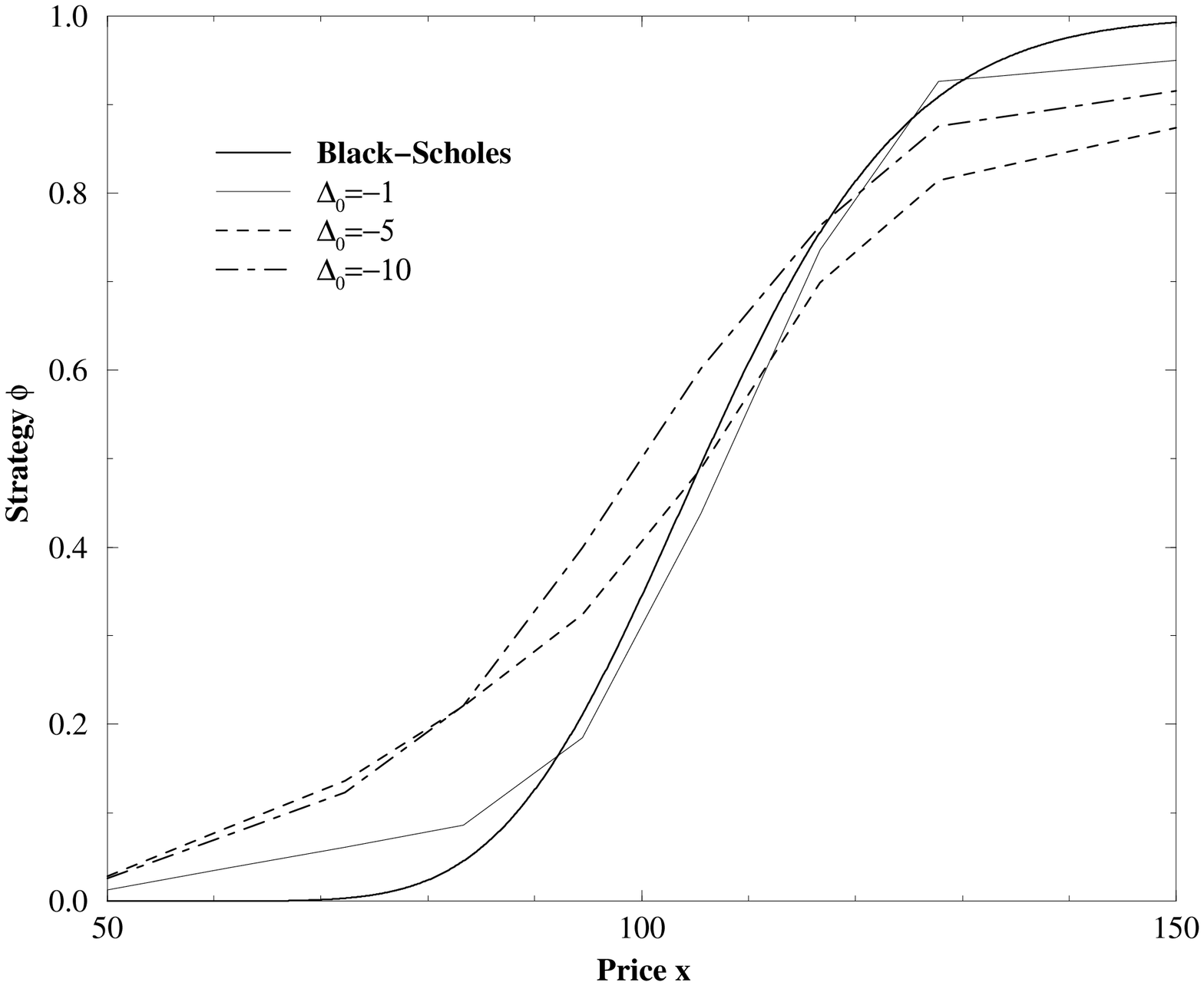}
    \caption{\label{finu} Optimal number of risky assets $\phi$ in the hedging portfolio,
    as a function of the level $x$ of the underlying asset for different strategies
    in the case of a fat tailed market (left: $\nu=6$ and right: $\nu=4$).
    The hedging strategies now present a very different dependence on the underlying, in
    particular when $|\Delta_0|$ increases. The value of the strike is $K=110$ and the time is half the maturity of the
        option: $k=N/2$.}
  \end{center}
\end{figure}

We are now interested in a market whose dynamics is governed by a
fat tailed L\'evy process, where the relative price increments are
independent identically distributed with jumps. In this case the
market is no longer complete (existence of unhedgeable jumps) and
perfect replication does not exist. The use of a subjective
criteria is then needed for pricing and hedging purposes, we now
expect to get different strategies depending on the value of
$\Delta_0$. Our expectations are numerically confirmed in Fig.
\ref{finu} where we clearly see strong differences between the
strategies. In particular, we observe that extreme losses hedging
(corresponding to a large value of $|\Delta_0|$) leads to a
flatter function $\phi(x)$ (i.e. a smaller Gamma). This was
already emphasized in \cite{Selmi,Potters} and can be quite
interesting in presence of transaction costs (see
section 4).

Using an independent set of paths, we can check that our method
indeed leads to smaller local expected shortfalls when the
corresponding optimal strategy is adopted, as can be seen in
Tables \ref{esf6} and \ref{esf4}, where we give both the
unconditional expected shortfall $\mc{R}$, and the conditional
expected shortfall (noted ESF), defined as $\mc{R}/{\cal P}$,
where ${\cal P}$ is the probability to exceed the threshold
$\Delta_0$. These expected shortfalls are computed between two
re-hedging times $k$ and $k+1$, where $k=N/2$ corresponds to half
the life of the option.

\begin{table}[!h]
\begin{center}
\begin{tabular}{|c|c|c|c|c|} \hline
\textbf{strategy}&BS&$\Delta_0=0$ & $\Delta_0=-5$ & $\Delta_0=-10$\\
\hline ESF(0)&-0.77&-0.88&-0.93&-1.19 \\
\hline ESF(-5)&-2.26&-2.32&-2.17&-1.98 \\
\hline ESF(-10)&-4.38&-4.00&-3.05&-2.21\\
\hline $\mc{R}(0)$&-0.22&-0.25&-0.33&-0.44 \\
\hline $\mc{R}(-5)$&-0.011&-0.014&-0.010&-0.015 \\
\hline $\mc{R}(-10)$&-0.002&-0.003&-0.002&-0.001\\
\hline
\end{tabular}
\end{center}
\caption{\label{esf6}{\small Conditional expected shortfall (ESF) and
unconditional expected shortfall $\mc{R}$, for different threshold
$\Delta_0$ and different strategies, in the case $\nu=6$. These
{\it local} quantities are computed for a time equal to half the
maturity of the option (here 1 year), for a strike $K=110$.}}
\end{table}

\begin{table}[!h]
\begin{center}
\begin{tabular}{|c|c|c|c|c|} \hline
\textbf{strategy}&BS&$\Delta_0=0$ & $\Delta_0=-5$ & $\Delta_0=-10$\\
\hline ESF(0)&-0.88&-1.04&-1.04&-1.10 \\
\hline ESF(-5)&-4.45&-4.40&-3.94&-3.15 \\
\hline ESF(-10)&-8.58&-8.60&-7.76&-5.90\\
\hline $\mc{R}(0)$&-0.23&-0.25&-0.35&-0.38 \\
\hline $\mc{R}(-5)$&-0.031&-0.035&-0.028&-0.028 \\
\hline $\mc{R}(-10)$&-0.015&-0.016&-0.011&-0.009\\
\hline
\end{tabular}
\end{center}
\caption{\label{esf4}{\small  Conditional expected shortfall (ESF) and
unconditional expected shortfall $\mc{R}$, for different threshold
$\Delta_0$ and different strategies, in the case $\nu=4$. These
{\it local} quantities are computed for a time equal to half the
maturity of the option (here 1 year), for a strike $K=110$.}}
\end{table}

\begin{figure}[!htbp]
  \begin{center}
    \includegraphics[angle=0,scale=0.4]{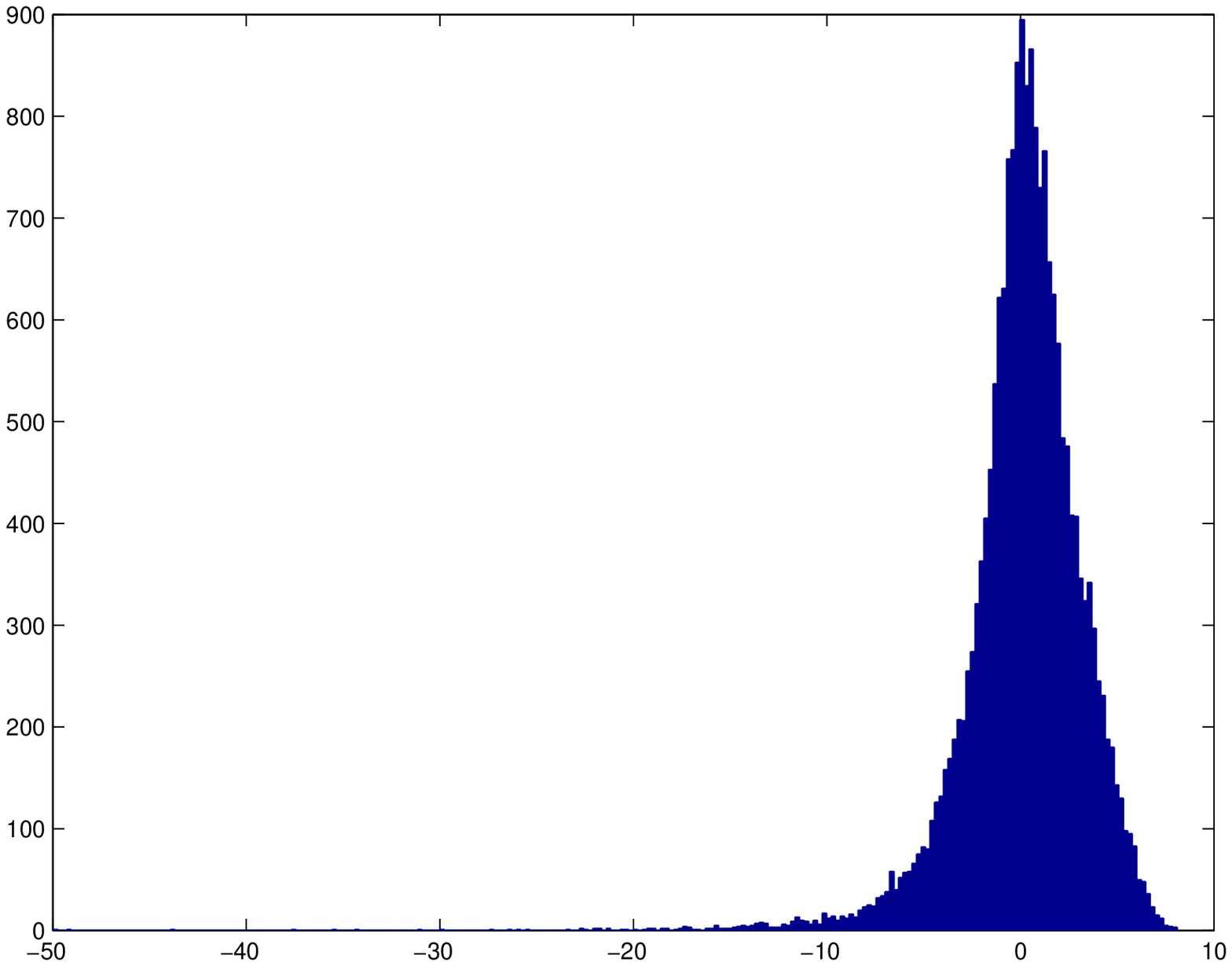}\includegraphics[angle=0,scale=0.4]{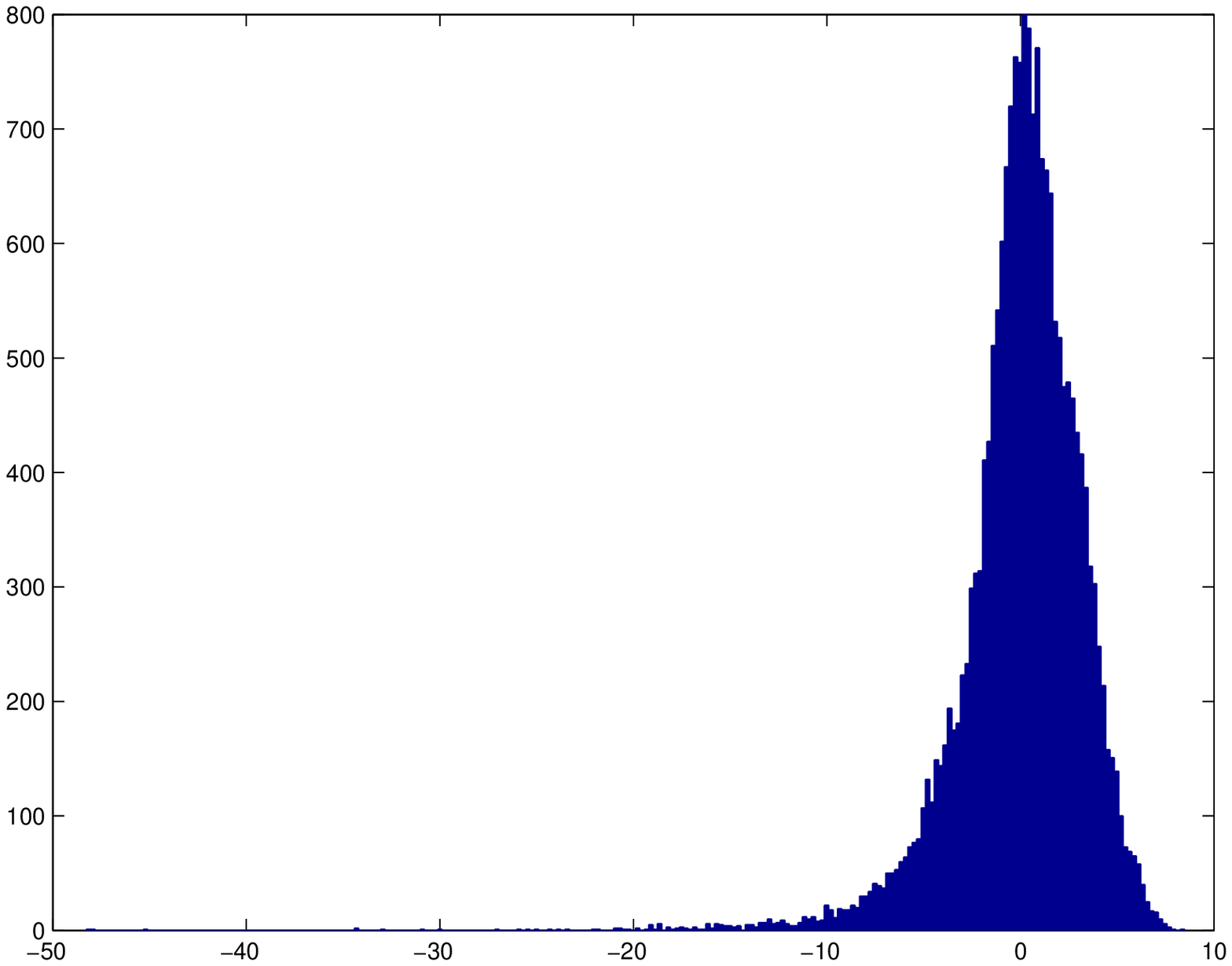}
    \includegraphics[angle=0,scale=0.4]{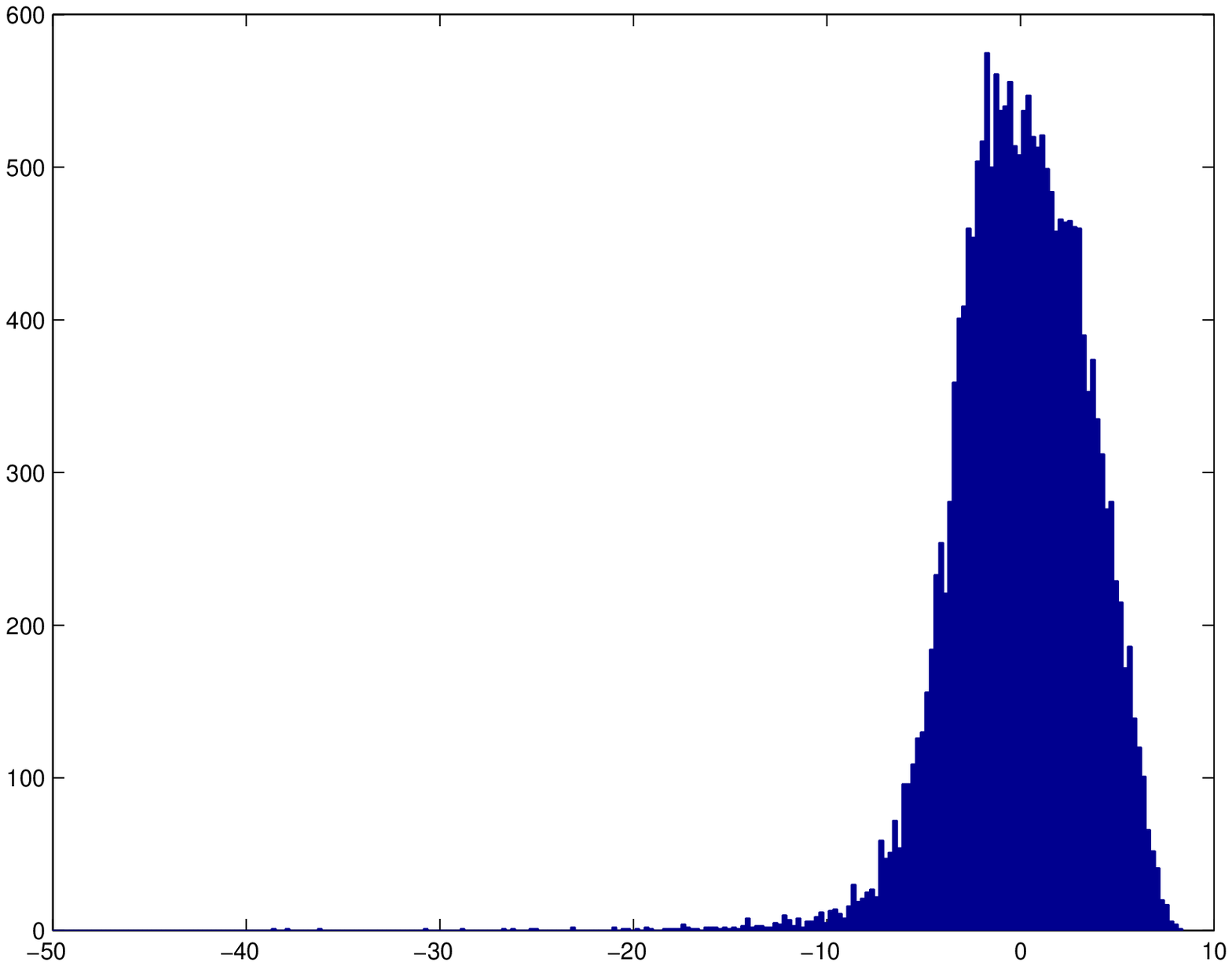}\includegraphics[angle=0,scale=0.4]{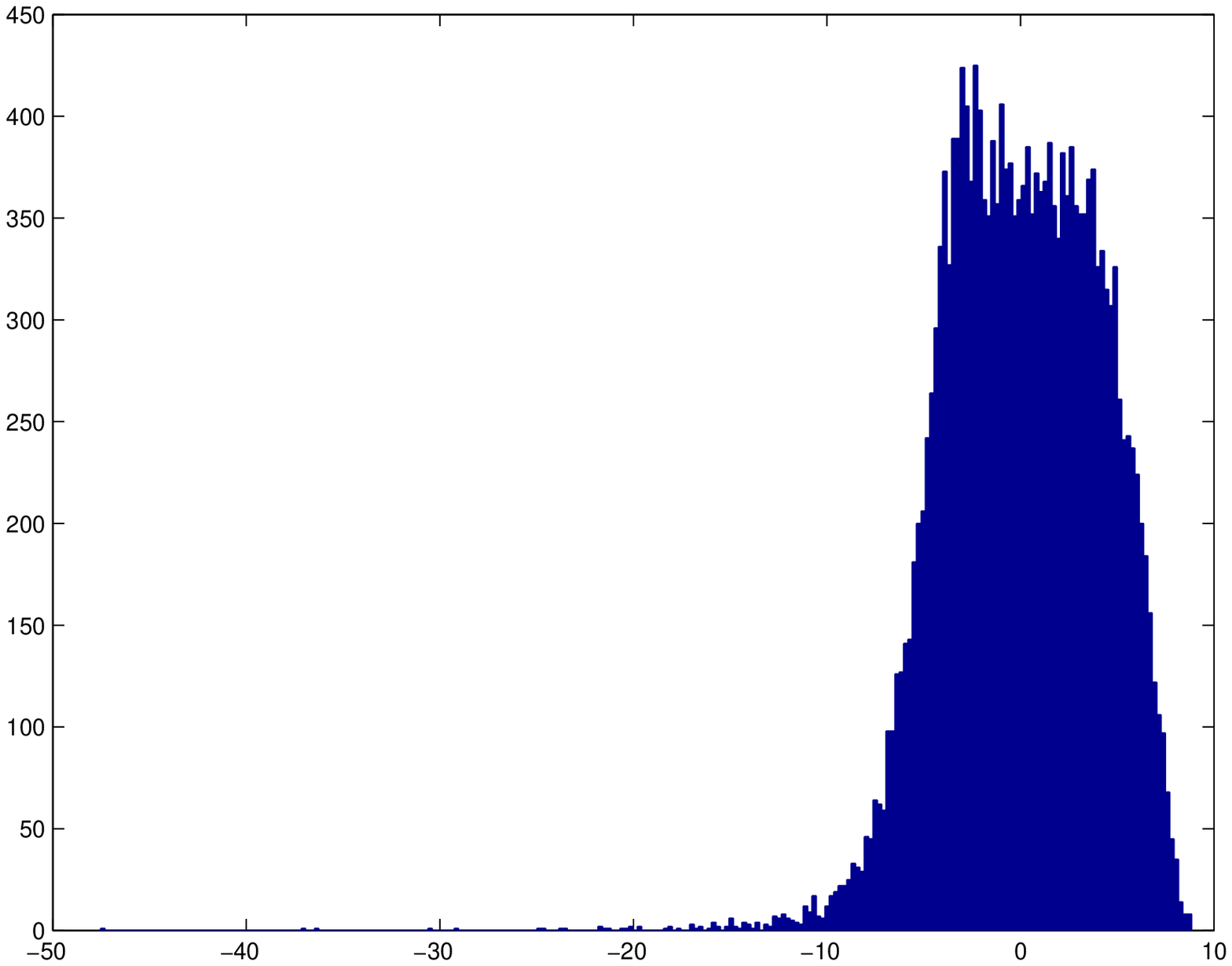}
    \caption{\label{d6f}  Distribution of the final wealth for different
    strategies in the case of a fat-tailed market with $\nu=6$
    (upper left: BS, upper right: $\Delta_0=0$, lower left: $\Delta_0=-5$, lower right:
    $\Delta_0=-10$).}
  \end{center}
\end{figure}

\begin{figure}[!htbp]
  \begin{center}
    \includegraphics[angle=0,scale=0.4]{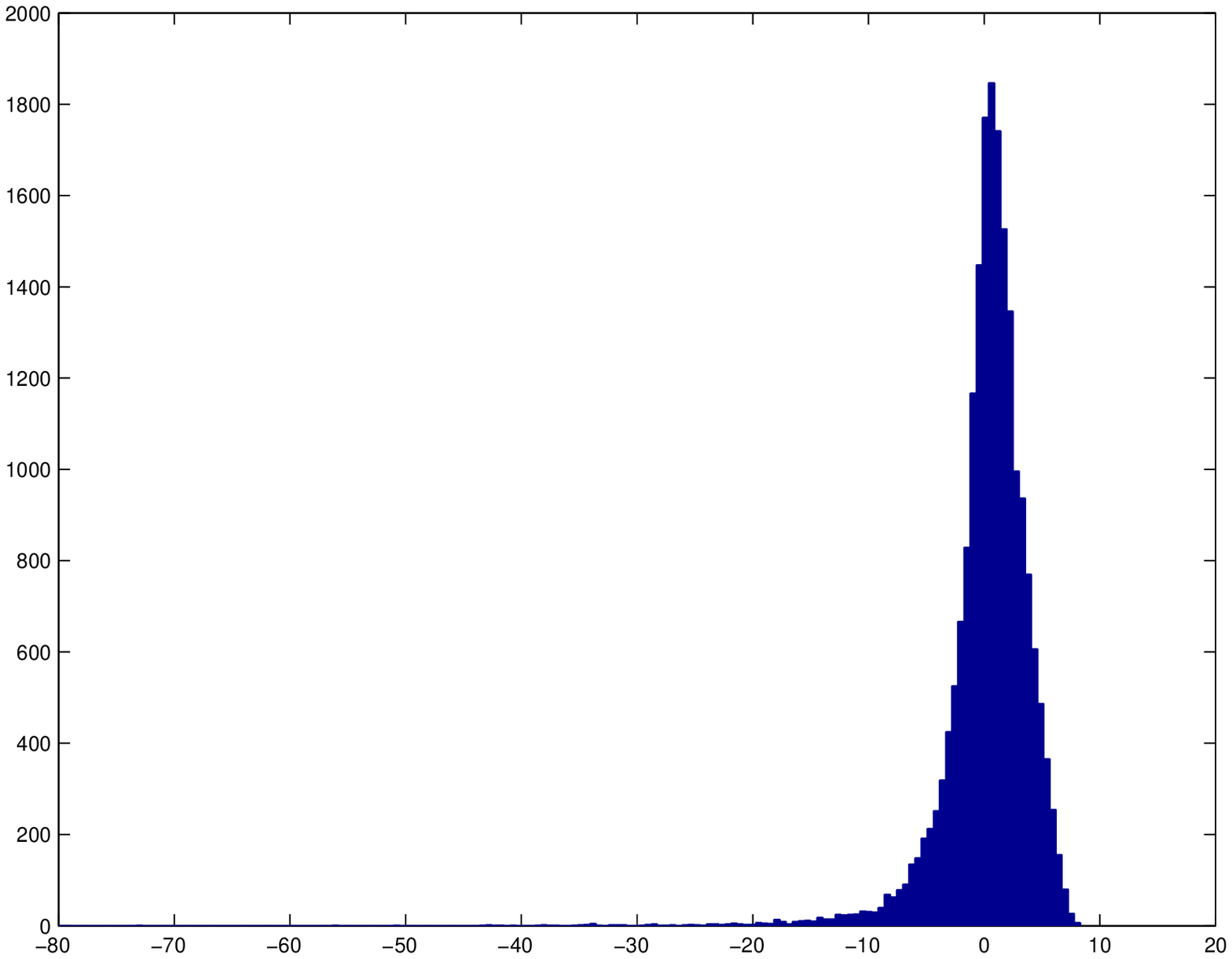}\includegraphics[angle=0,scale=0.4]{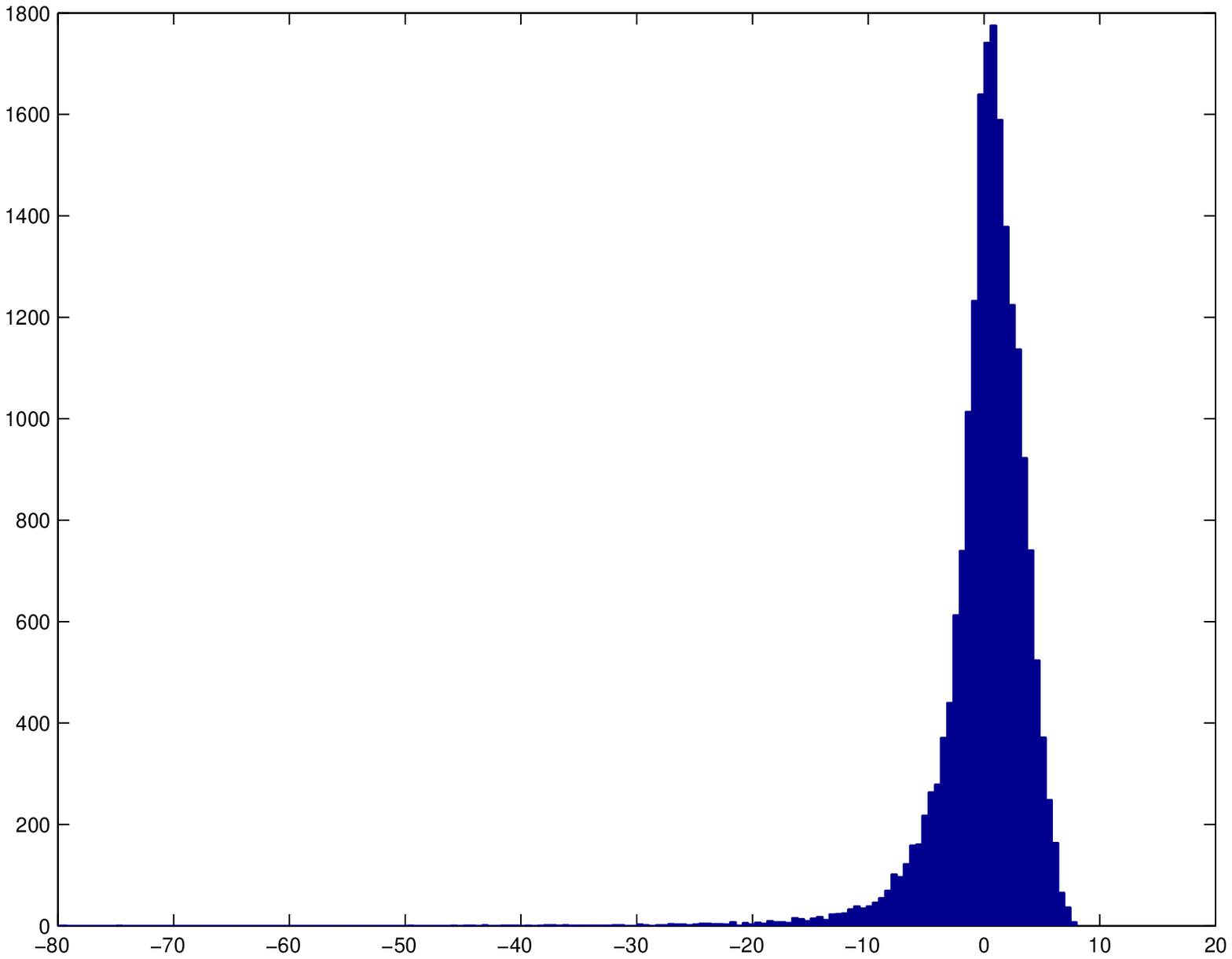}
    \includegraphics[angle=0,scale=0.4]{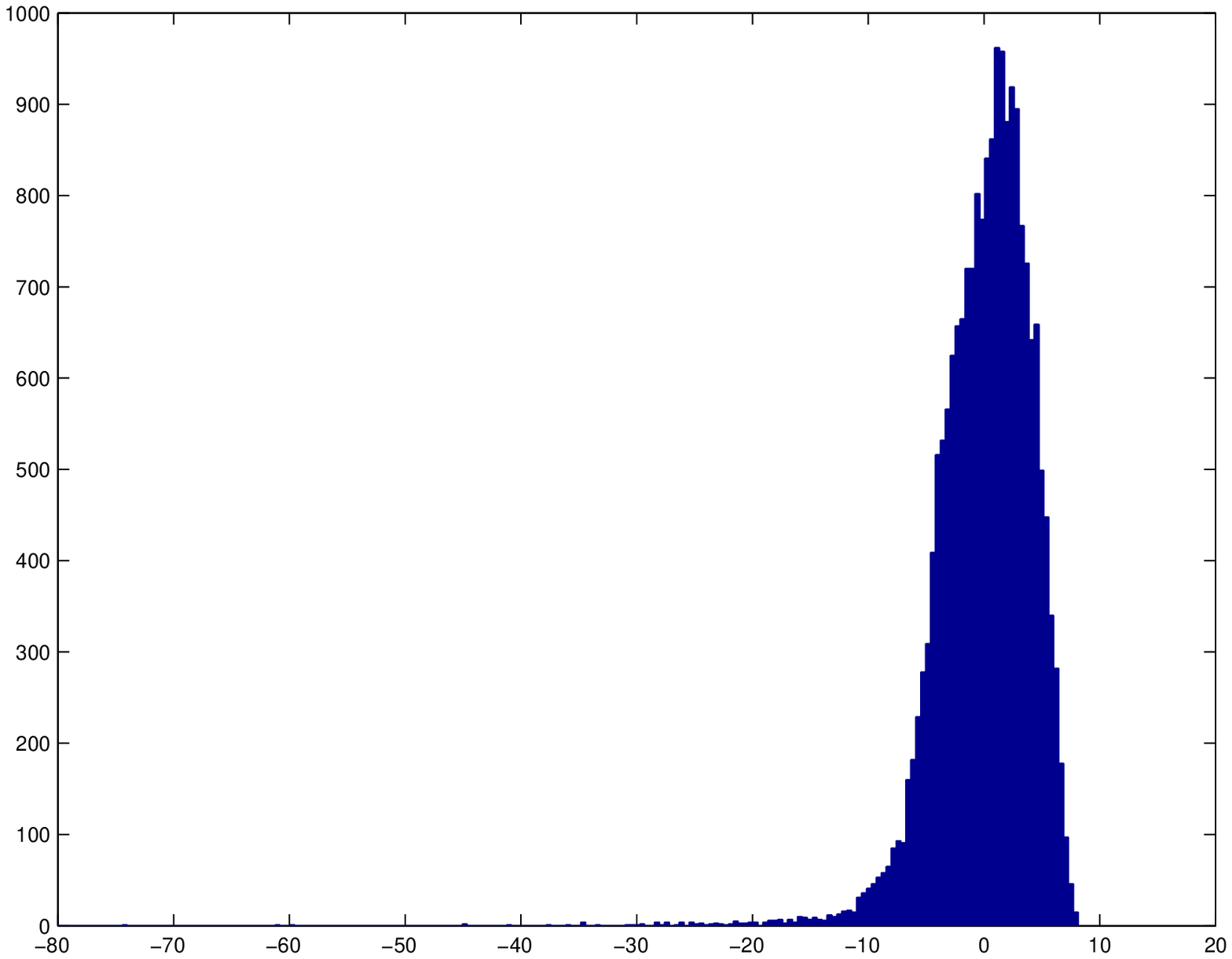}\includegraphics[angle=0,scale=0.4]{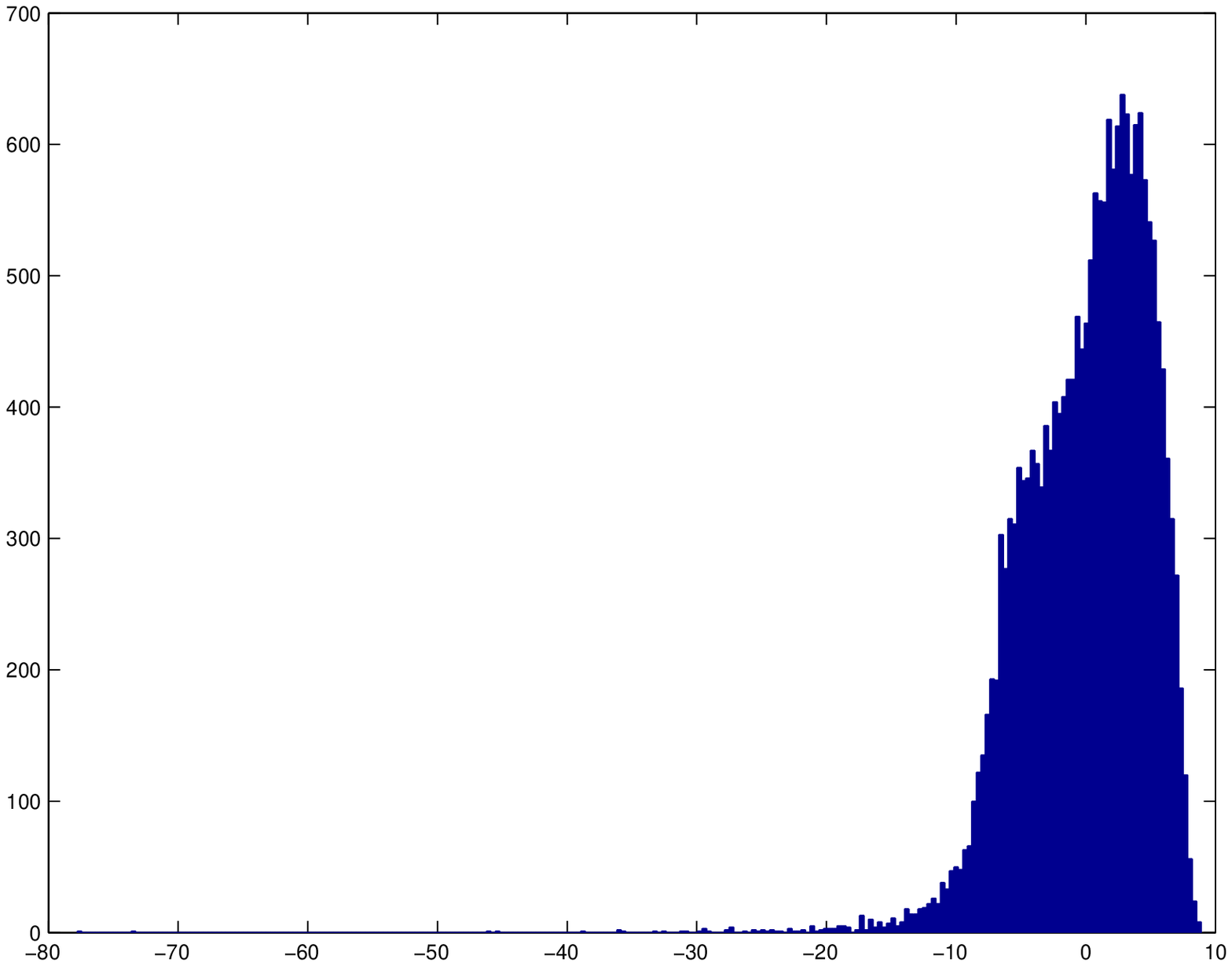}
    \caption{\label{d4f} distribution of the final wealth for different
strategies in the case of a fat-tailed market with $\nu=4$
    (upper left: BS, upper right: $\Delta_0=0$, lower left: $\Delta_0=-5$,
    lower right: $\Delta_0=-10$).}
  \end{center}
\end{figure}

We can also check that our method leads to satisfactory results
for {\it global} quantities (i.e. concerning the wealth balance at
the end of the option lifetime). We therefore determined the
distribution of the final wealth (Fig. \ref{d6f} and \ref{d4f}).
We clearly see that the strategy proposed here can significantly reduce the
value of the extreme losses (note that because of the power-tails
of the return distribution, these extreme losses can still be
large). Moreover we observe a remarkable change in the shape of
this distribution: from relatively peaked for small $|\Delta_0|$
but with an appreciable number of extreme events, it becomes
broader but with a smaller support when $|\Delta_0|$ increases.
This fact can be qualitatively explained: when $|\Delta_0|$ is
large, the constraint is easier to fulfill but losses of amplitude
less than $|\Delta_0|$  are not penalized, leading to a broader
looking but more sharply truncated final distribution.

\begin{figure}[!htbp]
  \begin{center}
    \includegraphics[angle=0,scale=0.75]{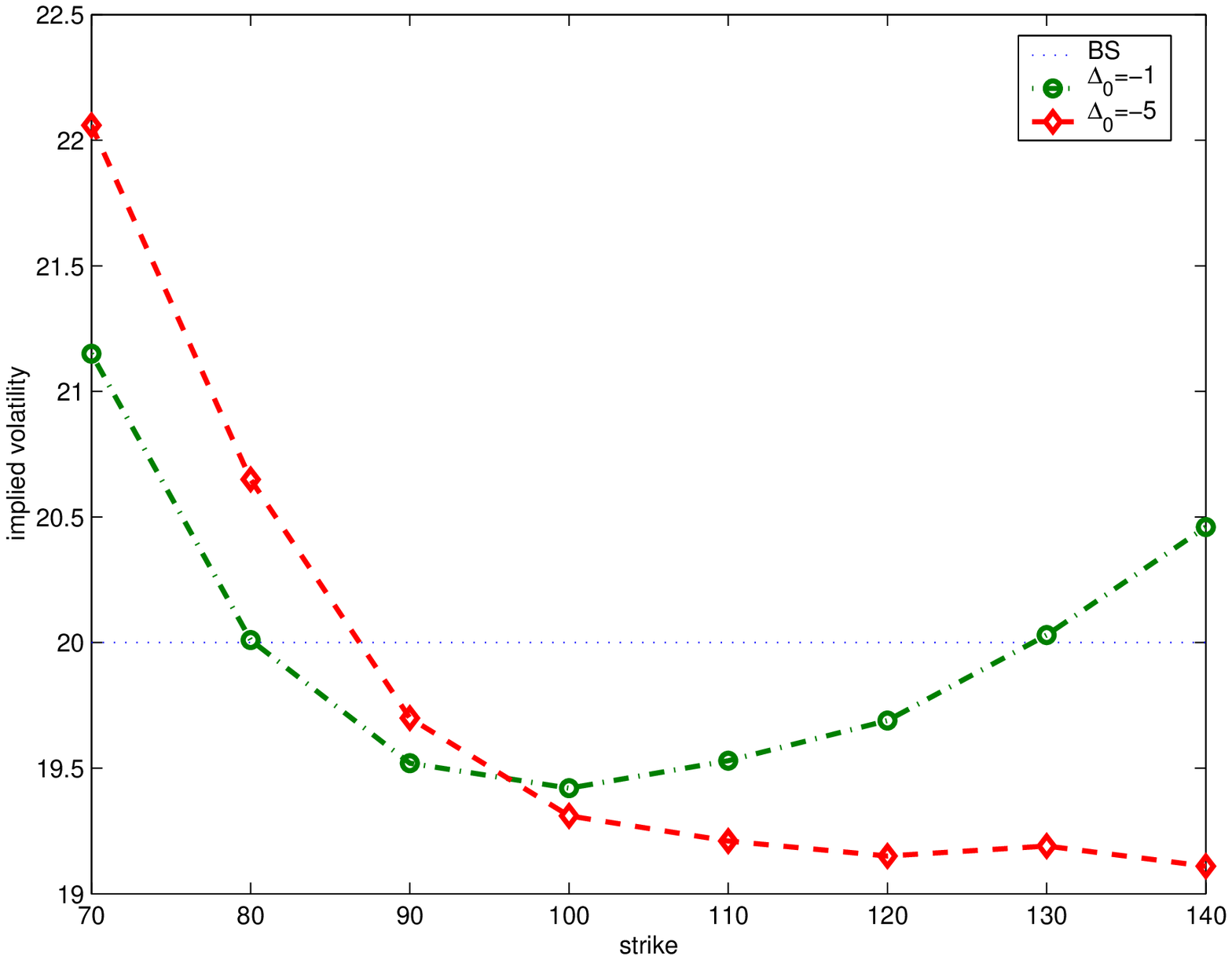}
    \caption{\label{smile} implied volatility smile at $t=0$. $\nu=4$ and
    $S_0=100$
    }
  \end{center}
\end{figure}

We give in Tables \ref{d6t} and \ref{d4t} the mean and the
standard deviation of these final wealth distributions, as well as
the initial price of the option and the associated value at risk
and expected shortfalls. Notice that the option price is smaller
than the Black-Scholes price, which is expected when the moneyness
is small: non zero kurtosis indeed leads to a {\it decrease} of
the at-the-money volatility \cite{Potters} as can be seen in Fig.
\ref{smile}. The option price in fact decreases when larger risks
are hedged.

\begin{table}[!h]
\begin{center}
\begin{tabular}{|c|c|c|c|c|} \hline
\textbf{strategy} & BS & $\Delta_0=0$ & $\Delta_0=-5$ & $\Delta_0=-10$\\
\hline \hline option price & 5.29 & 5.15 & 5.14 & 5.08 \\
\hline mean of final wealth &0.15 & 0.01 & 0.03 &0.04 \\
\hline std of final wealth &3.11 & 3.23 & 3.41 &4.01 \\
\hline VaR $0.1\%$ &-21.81 & -20.77 & -19.04 &-18.40 \\
\hline ESF $0.1\%$ & -7.96&-8.50&-7.63&-6.73 \\
\hline VaR $1\%$ &-10.04&-10.63&-8.82&-9.47\\
\hline ESF $1\%$ &-4.88&-4.91&-4.41&-3.69\\
\hline VaR $5\%$ &-4.85&-5.39&-5.28&-6.19\\
\hline ESF $5\%$ &-3.39&-3.46&-2.56&-2.28\\
\hline
\end{tabular}
\end{center}
\caption{\label{d6t} Statistical characteristics of the global
wealth balance for $\nu=6$. The strike is $K=110$. }
\end{table}

\begin{table}[!h]
\begin{center}
\begin{tabular}{|c|c|c|c|c|} \hline
\textbf{strategy} & BS & $\Delta_0=0$ & $\Delta_0=-5$ & $\Delta_0=-10$\\
\hline \hline option price & 5.29 & 5.08 & 4.99 &4.89 \\
\hline mean of final wealth &0.29 & 0.09 & 0.07 &0.09 \\
\hline std of final wealth &4.02 & 4.12 & 4.20 &4.82 \\
\hline VaR $0.1\%$ &-34.90 & -36.49 & -28.37 &-27.45 \\
\hline ESF $0.1\%$ & -16.14&-15.21&-14.05&-9.87 \\
\hline VaR $1\%$ &-13.59&-14.28&-12.08&-12.72\\
\hline ESF $1\%$ &-9.34&-9.34&-7.85&-6.06\\
\hline VaR $5\%$ &-5.58&-6.19&-6.32&-7.76\\
\hline ESF $5\%$ &-5.48&-5.55&-4.23&-3.51\\
\hline
\end{tabular}
\end{center}
\caption{\label{d4t} Statistical characteristics of the global
wealth balance for $\nu=4$. The strike is $K=110$.}
\end{table}

\section{Application to transaction costs}

As mentioned above, hedging against extreme risks generically leads to a strategy
that varies more slowly with the underlying asset price. This can be of great interest
in the presence of transaction costs. These costs
can in fact be endogenously taken into account within
the present numerical scheme, which allows to determine how both the price and the
optimal hedge are impacted by transaction costs. Previous analytical work on this
problem in the framework of the BS model can be found in \cite{Leland,Wilmott}, and
further discussions and extensions to the case of non Gaussian markets can be found in
\cite{Selmithese}.

\begin{figure}[!htbp]
  \begin{center}
    \includegraphics[angle=0,scale=0.55]{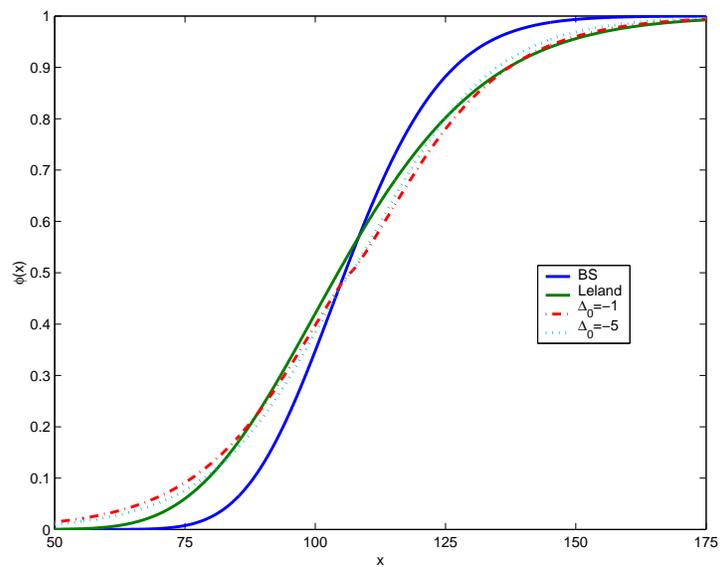}
    \caption{\label{ficost} Optimal number of risky assets $\phi$ in the hedging portfolio,
    as a function of the level $x$ of the underlying asset for different strategies:
        Black-Scholes, Leland, $\Delta_0=-1$ and $\Delta_0=-5$, for $\beta=0.05$.
    The value of the strike is K=$110$.
    }
  \end{center}
\end{figure}

We model friction by adding to the wealth balance a cost
proportional to the number of bought or sold assets and to its
price. Eq.\ref{balance} now becomes
\begin{equation}\label{balance cost}
   \Delta W_k=e^{\rho}\mc{C}_k(x_k)-
   \mc{C}_{k+1}(x_{k+1})+\phi_k(x_k)(x_{k+1}-e^{\rho} x_k)-\beta x_k |\phi_k-\phi_{k-1}|,
\end{equation}
where $\beta x_k$ represents the transaction costs per share.
Following the same steps as above, we now want to minimize the risk function
\begin{eqnarray*}
    \mc{R^*}_k & = & \sum_{\ell=1}^{N_{MC}}
    \bigg (\Delta_0 -\left
    (e^{\rho}C_{k+1}(x_k^\ell)-C_{k+1}(x_{k+1}^\ell)\right)-
    (x_{k+1}^\ell-e^{\rho} x_k^\ell) \phi_k(x_k^\ell)\\
    & &+\beta x_k^\ell |x_{k}^\ell-x_{k-1}^\ell|\frac{\partial
\phi}{\partial x}(x_{k}^\ell)
    \bigg)_+,
    \end{eqnarray*}
where we approximated $|\phi_k-\phi_{k-1}|$ by $\frac{\partial
\phi}{\partial x}|\Delta x|$. This is justified if the time step
is sufficiently small, and is the key step to make the problem
tractable. However, since this involves the derivative of $\phi$, we
have preferred to work with a smooth parameterization of the function
$\phi_k$ with only two optimization parameters, rather than the full
decomposition over a set of basis functions, as was used above. We have checked
that the following choice gives very similar results than the ones obtained above
in the absence of transaction costs. We thus take:
$$\phi_k(x)=\frac{1}{2} \left(1+\left[\tanh \left|A_k {\cal M}_k
\right|\right]^{\beta_k}.\text{sign}({\cal M}_k)\right)
$$
with a rescaled moneyness ${\cal M}_k$ given by:
$$
{\cal M}_k= \frac{x-K
e^{-r(T-t_k)}}{\sigma \sqrt{T-t_k}}.
$$
By varying the two variational parameters $A_k$ and $\beta_k$, we
can, at each time step, optimize any risk measure. Using
essentially the same numerical optimization procedure as above we
are able to find the parameters $A_k$ and $\beta_k$, and thus the
optimal strategy $\phi_k^*$. In order to obtain the option price,
we have then to solve the following least square problem:

\begin{eqnarray*}
    \min_{\gamma} \,\,\, & &\sum_{\ell=1}^{N_{MC}} \left[\sum_{a=1}^{p}\right.
     \gamma_a^k C_a^k(x_k^\ell)-
      e^{-\rho} \bigg (\mc{C}_{k+1}(x_{k+1}^\ell)
      -\phi_k(x_k^\ell)(x_{k+1}^\ell-e^{\rho}
    x_k^\ell)\\
    &&\left.+\beta x_k^\ell |x_{k}^\ell-x_{k-1}^\ell|\frac{\partial
\phi}{\partial x}(x_{k}^\ell)\bigg)  \right]^2
    \end{eqnarray*}
We have numerically tested the above scheme in the case of a BS
market with the same characteristics as in the previous section,
and for different values of the friction parameter $\beta$. We
compared our results with those obtained following a naive BS
strategy and the more advanced Leland strategy. Using simple
arguments, similar in spirit to the above approximation on $\Delta
\phi$, Leland showed in \cite{Leland} how the BS strategy can be
modified to account for transaction costs. Indeed using a modified
volatility
$$
\sigma_L=\sigma \sqrt{1+2\beta\sqrt{\frac{2}{\pi}}\frac{1}{\sigma
\sqrt{\Delta t}}},
$$
instead of the real volatility $\sigma$ leads to a strategy which, on average,
approximately covers the transaction costs and hedges the risk. Since
$\sigma_L>\sigma$, the option price is, as expected, higher than
in the BS case. Figure \ref{ficost} compares the obtained optimal
strategies using our method with both the BS and Leland hedging schemes.
As expected, the costs affects the BS strategy in
such a way as to reduce its at the money Gamma. Tables
\ref{tcb}, \ref{tcl}, \ref{tc1} and \ref{tc2} give a summary
statistics of our results. We note that even in the case where the stock
price is log-normal, our strategy allows to improve significantly over the Leland
strategy if the threshold $|\Delta_0|$ is large enough: compare Tables
\ref{tcl} and \ref{tc2}. Using the optimal strategy allows one to simultaneously
reduce the occurrence of large risks (measures both by the VaR and the ESF) while
keeping the option price lower than in the Leland scheme.

\begin{table}[!h]
\begin{center}
\begin{tabular}{|c|c|c|c|} \hline
$\beta$ & 0.005 & 0.01 & 0.05\\
\hline \hline option price & 5.29 & 5.29& 5.29 \\
\hline mean of final wealth &-0.41&-0.90&-4.78 \\
\hline std of final wealth &2.38&2.45&3.28\\
\hline VaR $0.1\%$ &-10.95&-11.86&-19.91 \\
\hline ESF $0.1\%$ & -1.45&-1.47&-1.33 \\
\hline VaR $1\%$ &-7.32&-8.06&-14.83\\
\hline ESF $1\%$ &-1.57&-1.67&-2.07\\
\hline VaR $5\%$ &-4.47&-5.15&-11.14\\
\hline ESF $5\%$ &-1.75&-1.84&-2.25\\
\hline
\end{tabular}
\end{center}
\caption{\label{tcb} Impact of the transaction costs for the B\&S strategy. }
\end{table}

\begin{table}[!h]
\begin{center}
\begin{tabular}{|c|c|c|c|} \hline
$\beta$ & 0.005 & 0.01 & 0.05\\
\hline \hline option price & 5.69& 6.08& 9.27 \\
\hline mean of final wealth &0.10&0.13 &0.32\\
\hline std of final wealth &2.39&2.44&2.98\\
\hline VaR $0.1\%$ &-9.83&-9.79& -10.32\\
\hline ESF $0.1\%$ &-1.54&-1.53 &-1.28\\
\hline VaR $1\%$ &-6.47&-6.47&-6.83\\
\hline ESF $1\%$ &-1.55&-1.54&-1.60\\
\hline VaR $5\%$ &-3.80&-3.84&-4.27\\
\hline ESF $5\%$ &-1.67&-1.64&-1.62\\
\hline
\end{tabular}
\end{center}
\caption{\label{tcl} Impact of the transaction costs for the Leland strategy.}
\end{table}

\begin{table}[!h]
\begin{center}
\begin{tabular}{|c|c|c|c|} \hline
$\beta$ & 0.005 & 0.01&0.05\\
\hline \hline option price & 5.77& 6.22& 8.30 \\
\hline mean of final wealth &0.00&-0.04 &-0.26\\
\hline std of final wealth &2.39&2.44&3.09\\
\hline VaR $0.1\%$ &-11.05&-10.95& -10.87\\
\hline ESF $0.1\%$ & -1.34&-1.45 &-1.27\\
\hline VaR $1\%$ &-7.05&-7.13&-7.54\\
\hline ESF $1\%$ &-1.61&-1.54&-1.45\\
\hline VaR $5\%$ &-4.01&-4.11&-5.09\\
\hline ESF $5\%$ &-1.84&-1.80&-1.51\\
\hline
\end{tabular}

\end{center}
\caption{\label{tc1} Impact of the transaction costs using the
optimal strategy, where these costs are accounted for.
$\Delta_0=-1$}
\end{table}

\begin{table}[!h]
\begin{center}
\begin{tabular}{|c|c|c|c|} \hline
$\beta$ & 0.005 & 0.01&0.05\\
\hline \hline option price  & 5.63& 5.99& 8.44\\
\hline mean of final wealth &0.00&-0.05 &-0.75\\
\hline std of final wealth &2.66&2.74&3.39\\
\hline VaR $0.1\%$ &-8.92&-9.17& -11.95\\
\hline ESF $0.1\%$ & -1.17&-1.27 &-1.44\\
\hline VaR $1\%$ &-6.12&-6.37&-8.60\\
\hline ESF $1\%$ &-1.28&-1.31&-1.48\\
\hline VaR $5\%$ &-4.06&-4.25&-6.07\\
\hline ESF $5\%$ &-1.30&-1.33&-1.56\\
\hline
\end{tabular}

\end{center}
\caption{\label{tc2} Impact of the transaction costs using the
optimal strategy, where these costs are accounted for.
$\Delta_0=-5$}
\end{table}

\section{Conclusion}

In this paper we have extended the work of \cite{PBS} and proposed
a general numerical (Monte-Carlo) methodology for the pricing and hedging of options
when the market is incomplete, for an arbitrary risk criterion (chosen here to
be the expected shortfall) and in the presence of transaction costs.
We have shown that in the presence of fat-tails, our strategy allows to
significantly reduce extreme risks, and generically leads to low Gamma
hedging, as anticipated in \cite{Selmi,Potters}. Many other risk criteria
could be considered, in particular functions that give more weights to
extreme losses. We focused in this work on plain vanilla European options,
but (as shown in \cite{PBS}) the method is readily extended to a large family
of exotic options. Finally, we showed how our method allows to deal consistently
with transaction costs. When compared to the standard Leland hedging scheme, our
optimal strategy leads both to lower option prices and better hedging of
large risks, even in the simplest case of a log-normal market.

There are many extensions of the above method that would be worth
investigating, in particular the case where the underlying has a
stochastic volatility with some persistence, such as, for example,
the models studied in \cite{Heston,Yakov,Masoliver,Muzy,Pochart}.
In this case, both the price and the optimal hedge should
explicitly depend on the local value of the volatility, or of a
noisy estimate of this volatility. Other hedging instruments, like
options of different maturities, could in this case be included in
the local wealth balance to reduce the risk further. Another
interesting path is to consider the problem of hedging a whole
portfolio of options. Contrarily to the Black-Scholes case, where
the optimal strategy for the whole portfolio is the linear sum of
the individual hedges, extreme value hedges lead to a non linear
composition of the individual hedges.

\bibliography{biblio_ESF}
\bibliographystyle{unsrt}
\end{document}